\documentclass{iau} 
\usepackage{amsmath}
\usepackage{mathrsfs}
\usepackage{tikz}
\usepackage{ulem,xpatch}
\usepackage{graphicx}

\newcommand{\orcidauthorA}{0000-0003-1315-4984}

\title[QPOs from post-shock accretion column] %% give here short title %%
{QPOs from post-shock accretion column of strongly magnetized accreting white dwarfs}

\author[Prasanta Bera]   %% give here short author list %%
{Prasanta Bera$^1\orcidauthorA$
}

\affiliation{$^1$Mathematical Sciences and STAG Research Centre, University of Southampton, Southampton SO17 1BJ, U.K. \\ email: {\tt pbera.phy@gmail.com} \\[\affilskip]
}

\pubyear{2019}
\volume{357}  %% insert here IAU Symposium No.
\jname{White Dwarfs as probes of fundamental physics% and tracers of planetary, stellar & galactic evolution
}
\editors{A.C. Editor, B.D. Editor \& C.E. Editor, eds.}
\begin{document}

\maketitle

\begin{abstract}
An extended magnetosphere of a strongly magnetized accreting white dwarf (known as a polar) prevents the formation of an accretion disk and the matter is channelled to the magnetic pole(s). A few such sources show quasi-periodic oscillations in their optical light curves. These high-frequency oscillations are thought to be generated from the post-shock accretion column. The kinetic energy of the accretion flow is finally emitted from this post-shock region and the involved radiation processes decide the state of the matter. Here we study the structure and the dynamical properties of such accretion columns and compare the results with the observational characteristics.

\end{abstract}

\firstsection % if your document starts with a section,
\section{Introduction}

A set of polars (e.g. AN~UMa, VV~Puppis, V834~Cen) show quasi-periodic oscillations in their optical light curves in various observations \cite[(Middleditch 1982, Mouchet et al. 2017)]{Middleditch1982, Mouchet+2017}. The presence of cyclotron lines and Zeeman splitting in their line spectra indicates the signature of a strong magnetic field at the surface of the white dwarf. A strong magnetic field at the near surroundings of a white dwarf in a binary stellar system modifies the motion of the accretion flow. The ionized accretion flow gains kinetic energy as it moves closer to the white dwarf and follows the magnetic field lines within the magnetosphere of the compact star. The freely falling accretion matter impacts at the magnetic pole of the compact star after forming a shock. Along with the motion of the flow, the kinetic energy of the pre-shocked material is transformed into the thermal energy of the post-shock phase. This thermal energy is eventually radiated via some processes e.g. bremsstrahlung, emitted photons from the accelerated charged particles in a field of ions; and cyclotron, emission from the gyrating electrons about the magnetic field \cite{Chanmugam+Wagner1979}. The stellar parameters indicate the high energy X-ray photons generated via bremsstrahlung emission escape freely from the shocked matter but the optical photons cause the thermal radiation from the outer surface of the post-shock region.

The presence of rapid quasi-periodic oscillation in these systems indicates that they should originate from a region with a small length scale affected by high speed. Different studies show the characteristics of the emitted radiation from the post-shocked region \cite{Langer+1981}. The theoretical models indicate the possibilities of QPOs in both the X-ray and optical energy bands as they are generated from the same post-shock region. However, X-ray observations have not confirmed any QPOs to date \cite[(Imamura et al. 2000, Bonnet-Bidaud et al. 2015)]{Imamura+2000, Bonnet-Bidaud+2015}. Hence to develop a consistent model for the dynamics of the post-shock requires us to study the overall post-shock region considering the involved radiation processes. 

\section{Model setup}\label{S2}
\begin{figure}[b]
% \vspace*{-2.0 cm}
\begin{center}
 \includegraphics[width=0.45\linewidth]{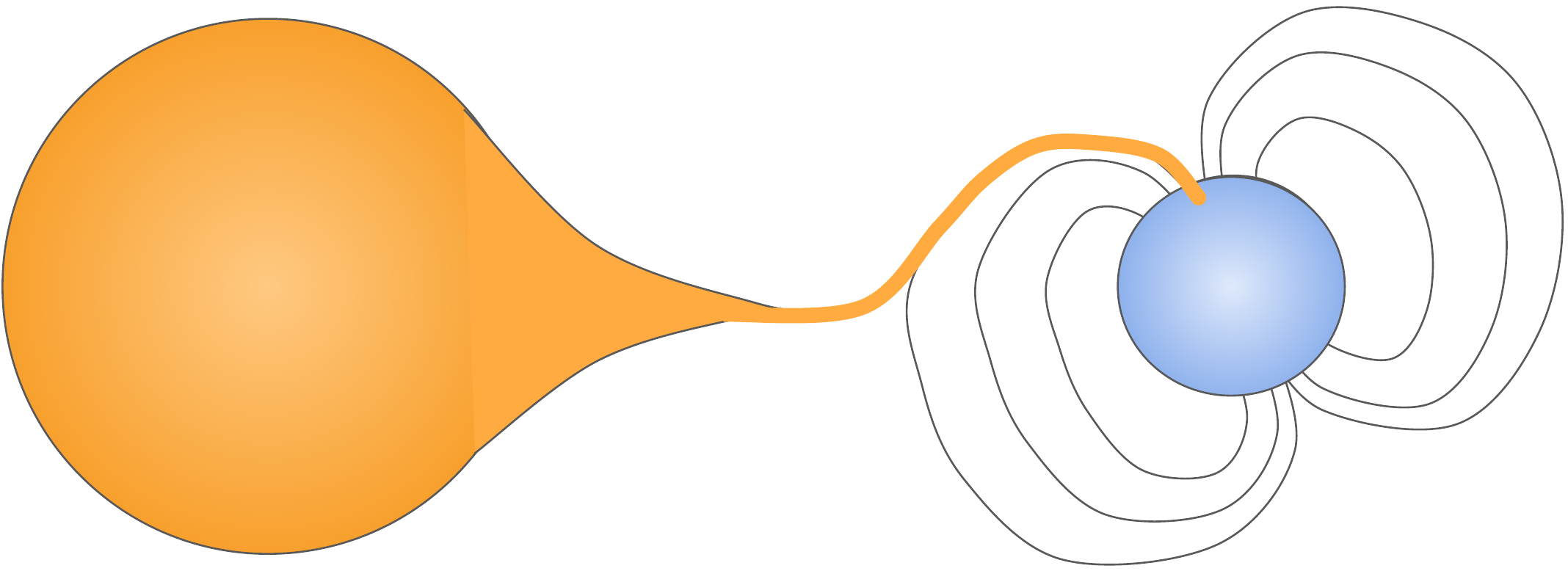} 
  \includegraphics[width=0.45\linewidth]{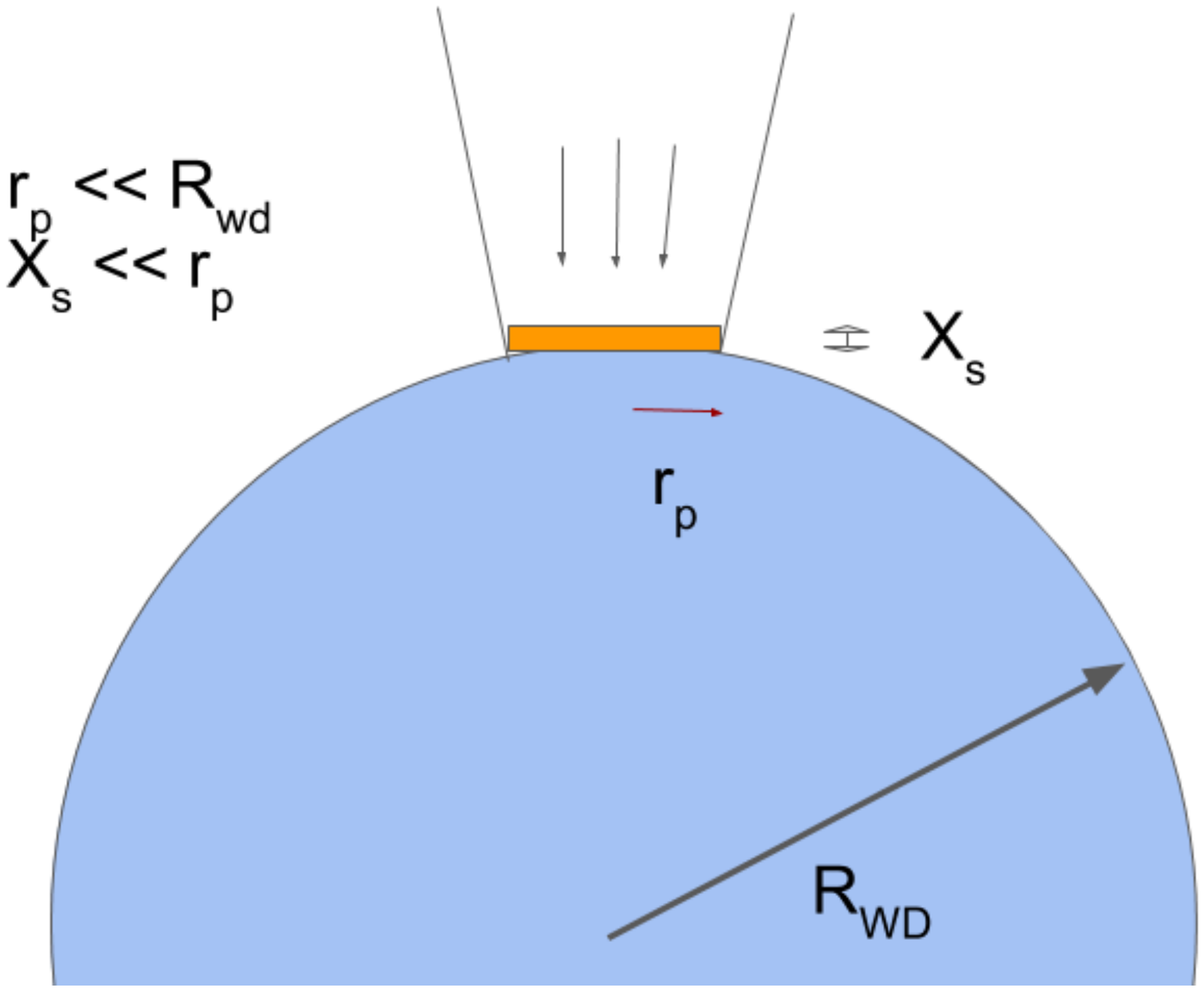}
% \vspace*{-1.0 cm}
 \caption{\textit{Left}: A cartoon image of a binary stellar system where one of the stars is a white dwarf (smaller in size) with a strong magnetic field. The other star overflows matter which is being accreted by the white dwarf at its magnetic pole.
 \textit{right}: The accretion column forms on the white dwarf surface at the magnetic pole. The distance of shock location from the surface $x_s$ is very small compared to the length of the cross-section of the accretion column.}
   \label{fig1}
\end{center}
\end{figure}

We consider matter accreted from the companion star to one of the magnetic poles of the white dwarf without the formation of an accretion disk (Fig.~\ref{fig1}). We further assume that at the magnetic pole, the field lines are vertical to the surface and we can exclude the local field geometry. The typical accretion parameters (e.g. inflow matter density, magnetic field strength) of the accreting polars suggest the height of the shock from the stellar surface is much less than the characteristic length scale of the cross-sectional area. In this condition, we study the vertical variations only.

The dynamics of accretion matter is described by the following hydrodynamic equations~\ref{continuty}, \ref{momentum_conservation} and \ref{energy_conservation} corresponding to the conservation of mass, momentum and energy respectively \cite{Hameury+1983}.
 \begin{align}
 \frac{\partial\rho}{\partial t} + \nabla\cdot(\rho\bf{v}) &= 0\label{continuty}\\
 \rho\left(\frac{\partial\mathbf{v}}{\partial t}+(\mathbf{v}\cdot\nabla)\mathbf{v}\right) &= -\mathbf{\nabla} p -\rho\mathbf{\nabla} \Phi \label{momentum_conservation}\\
 \frac{\partial(E+\rho\Phi)}{\partial t}+\nabla\cdot(\mathbf{v}(E+p+\rho\Phi)) &=  -\rho\mathbf{v\cdot\nabla} \Phi - \Lambda \label{energy_conservation}
\end{align}
here $\rho$: matter density, $\mathbf{v}$: velocity, $p$: pressure, $E=\frac{1}{2}\rho \mathbf{v\cdot v}+\rho\epsilon$: total energy density, $\epsilon=\frac{1}{\rho}\frac{p}{\gamma-1}$: internal energy per unit mass with adiabatic index $\gamma$, $\Lambda$ : cooling function, $\Phi$ : stellar gravitational potential, $t$: time and $\nabla$: spatial gradient. We solve these equations for the provided boundary conditions in one-spatial dimension i.e. along the vertical direction.

\section{Results}\label{S3}
\begin{figure}[b]
% \vspace*{-2.0 cm}
\begin{center}
 \includegraphics[width=0.7\linewidth]{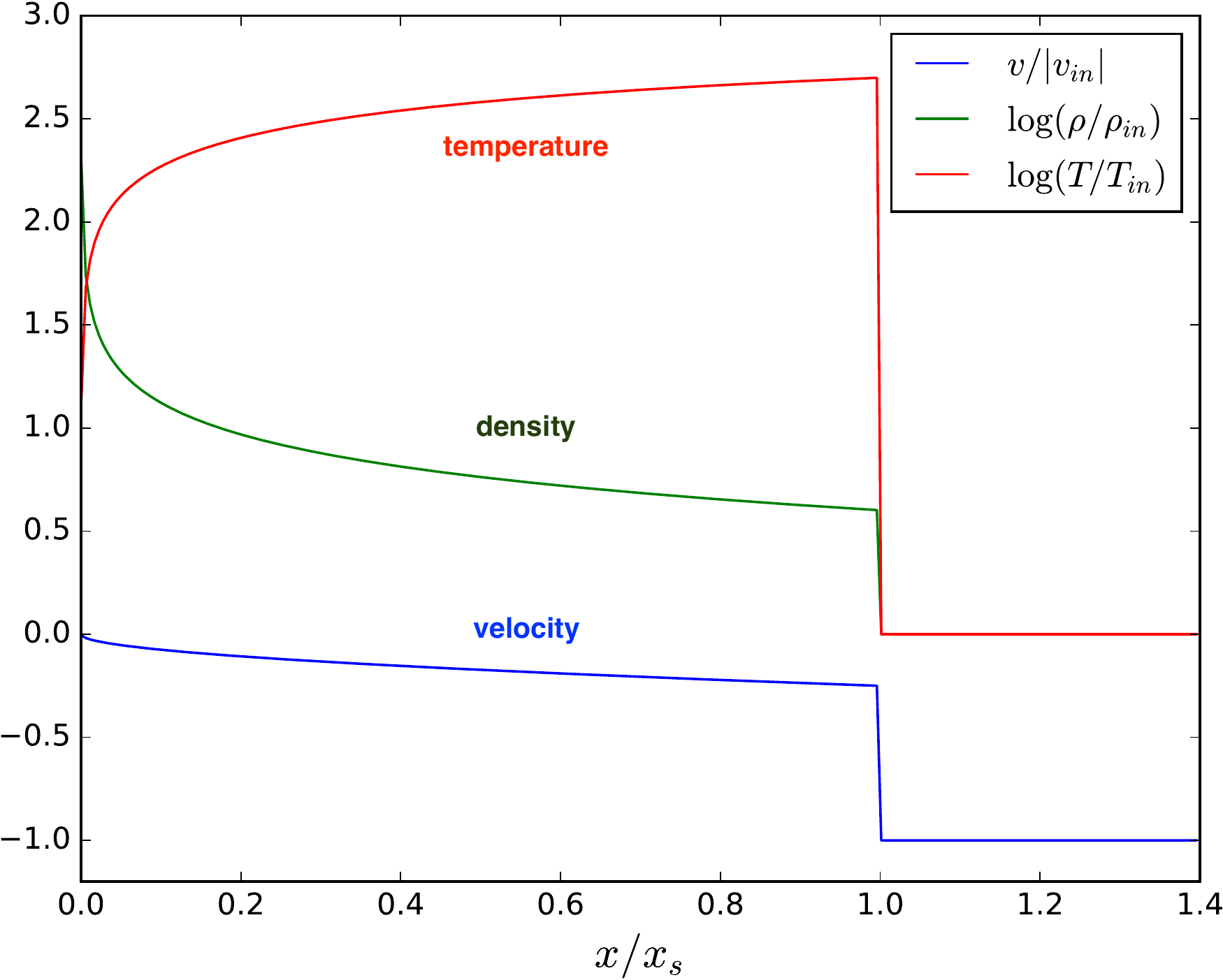} 
% \vspace*{-1.0 cm}
 \caption{The steady equilibrium solution of velocity, density and temperature at the post-shock region corresponding to a fixed accretion inflow and a similar mass absorption rate at the base. Here, bremsstrahlung is the only cooling process considered.}
   \label{fig2}
\end{center}
\end{figure}

In the steady accretion condition, for the provided boundary values the structure can be obtained from the solution of the above equations neglecting the time derivative terms. For a negligible inflow velocity at the base of accretion column, Fig.~\ref{fig2} shows the vertical variation of the flow velocity, density and temperature in terms of the corresponding inflow values. In the strong shock assumption, the jump in density has a discontinuity of a factor of four. The flow density increases towards the stellar surface but the temperature reduces to balance the pressure and the cooling processes. The bremsstrahlung process becomes significant at the base of the column where matter density is higher as $\Lambda_{brem}\sim \rho^2T^{1/2}$. The temperature-dependent cyclotron cooling ($\Lambda_{cycl}\sim \rho^{3/20}T^{5/2}$) becomes dominating just below the shock. Hence, the relative importance of the cooling processes decides the dynamics of the post-shock region \cite[(Bera, \& Bhattacharya 2018)]{Bera+Bhattacharya2018}.

\begin{figure}[b]
% \vspace*{-2.0 cm}
\begin{center}
 \includegraphics[width=0.65\linewidth]{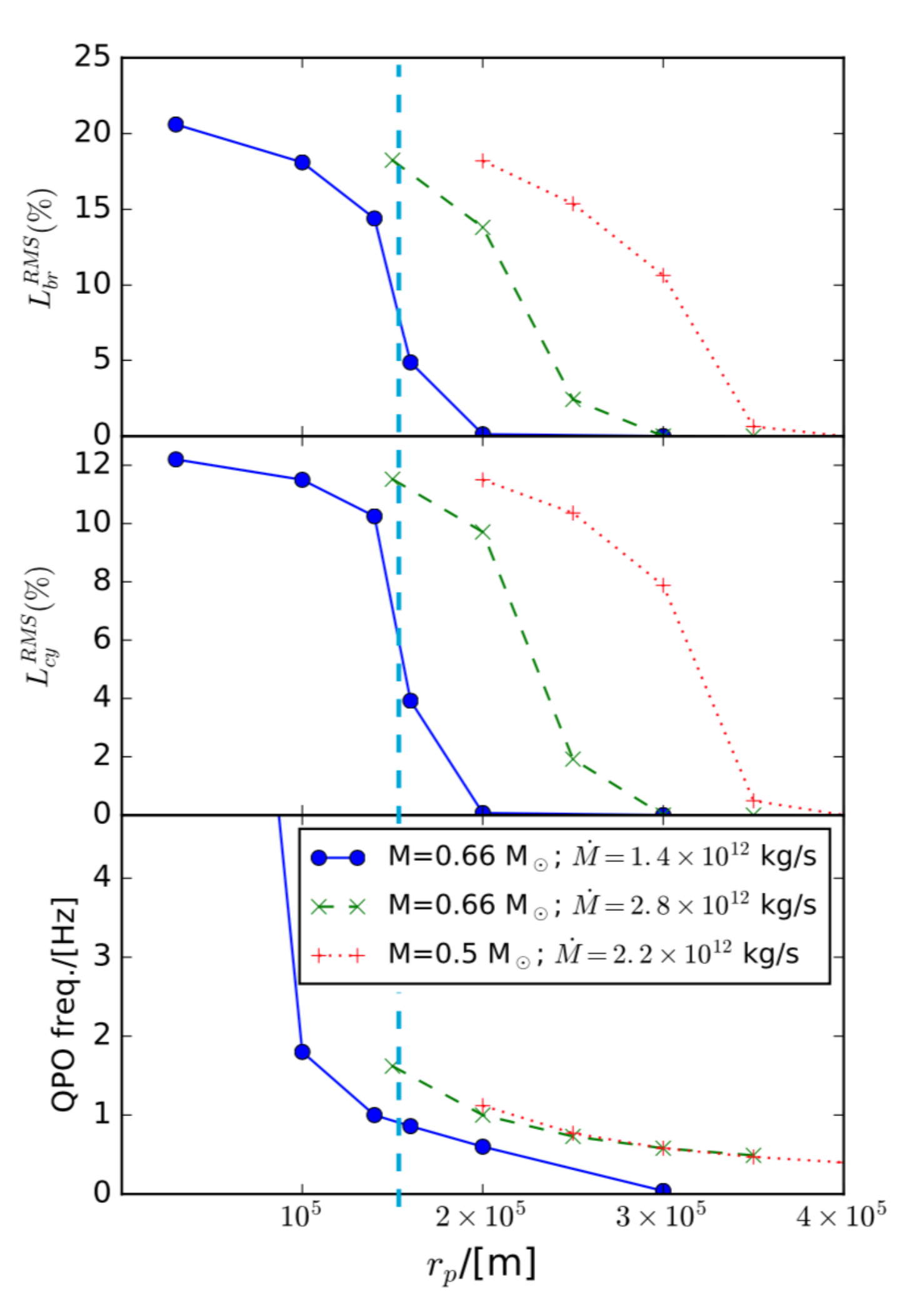} 
% \vspace*{-1.0 cm}
 \caption{Expected rms variability in bremsstrahlung, cyclotron radiation and corresponding QPO frequency from a dynamical evolution study of the accretion column for different polar cap radii. The accretion parameters are similar to the source V834~Cen. The results corresponding to $r_p=1.5\times10^5~\rm{m}$ on the magnetic pole of a $0.66~M_\odot$ mass white dwarf with the accretion rate $1.4\times10^{12}~\rm{kg/s}$ match well with the observed values.}
   \label{fig3}
\end{center}
\end{figure}

The temporal variability time scale is dependent on the cooling length and the inflow speed i.e. the characteristic cooling time scale. A perturbation of the steady solution indicates the characteristic oscillation time scale. We calculate the oscillation time scale from linear perturbation study and verify it from non-linear evolution \cite[(Mignone 2005, Bera \& Bhattacharya 2018)]{Mignone2005, Bera+Bhattacharya2018}. The non-linear evolution case provides the oscillation amplitude that is used to compare with the observational characteristics. Fig.~\ref{fig3} shows the expected model-dependent bremsstrahlung and cyclotron rms amplitudes and QPO frequencies for different cross-sectional size of the accretion column. The stellar parameters are used similar to those for V834~Cen and the results show consistency with observations.

\section{Summary}\label{S4}

The vertical structure and the dynamical properties of the post-shock accretion column formed on the surface of a magnetic white dwarf are studied here. We found that the characteristic time of oscillation is connected to the involved radiation processes. A lower accretion rate or a larger surface coverage of the accretion column reduces QPO frequency. The rms amplitude of the oscillation is maximum when bremsstrahlung is the only dominating cooling process.

\begin{acknowledgement}
I acknowledge the financial support from the organizers in terms of a travel grant to present this work at IAUS~357, Hilo.
\end{acknowledgement}

\def\mnras{MNRAS}%
\def\apj{ApJ}%
\def\apjl{ApJL}%
\def\aap{A\&A}% 
\def\pasp{PASP}%

\end{document}